\documentclass[11pt,a4paper]{article}

\pdfoutput=1
\usepackage{jheppub}

\usepackage{amssymb}
\usepackage{amsmath}
\usepackage{amsfonts}
\usepackage{graphicx}
\usepackage{color}
\usepackage{xspace}
\usepackage{ulem}
\usepackage{mathtools}
\usepackage{hhline}
\usepackage{float}
\usepackage{caption}
\usepackage{graphicx}

\begin{document}

\title{Determining the spin of light primordial black holes with Hawking radiation II: high spin regime}

\author[a]{Marco Calz\`a}    \emailAdd{mc@student.uc.pt}
\author[a]{Jo\~{a}o G.~Rosa} \emailAdd{jgrosa@uc.pt}

\affiliation[a]{Univ Coimbra, Faculdade de Ci\^encias e Tecnologia da Universidade de Coimbra and CFisUC, Rua Larga, 3004-516 Coimbra, Portugal}

\abstract{
   We propose a method to determine the mass and spin of primordial black holes based on measuring the energy and emission rate at the dipolar and quadrupolar peaks in the primary photon Hawking spectrum, applicable for dimensionless spin parameters $\tilde{a}\gtrsim 0.6$. In particular, we show that the ratio between the energies of the two peaks is only a function of the black hole spin, while the ratio between their emission rates depends also on the line-of-sight inclination. The black hole mass and distance from the Earth may then be inferred from the absolute values of the peak energies and emission rates. This method is relevant for primordial black holes born with large spin parameters that are presently still in the early stages of their evaporation process.}

\maketitle

\section{Introduction}

In 1975 Hawking showed that a black hole (BH) exhibits a nearly thermal emission spectrum characterized by a temperature inversely proportional to its mass, which in the case of a Kerr BH reads \cite{Hawking:1975vcx} 
\begin{equation}\label{THaw}
    T_{H}=   \frac{1  }{ (2+(\sqrt{1-\tilde{a}})^{-1})}{M_P^2\over 8\pi M}\;,
\end{equation}
where $\tilde{a}=J M_P^2/M^2$ is the dimensionless spin parameter, $M_P\sim 2.176 \times 10^{-8}$ kg is the Planck mass and $M$ and $J$ denote the BH mass and angular momentum, respectively. While this temperature is too low for stellar or supermassive BHs to yield any observable effects, it may be possible to detect the Hawking radiation emitted by light primordial black holes (PBHs) potentially formed in the early Universe \cite{Boluna:2023jwq,Coogan:2020tuf,Keith:2022sow,Agashe:2022phd,Khlopov:2008qy}. In fact, several observatories such as Fermi-LAT and Hawk have already been able to place constraints on the abundance of such compact objects, and several planned gamma-ray telescopes like AMEGO and ASTROGAM \cite{e-ASTROGAM:2016bph, Tatischeff:2019mun,AMEGO:2019gny, Ray:2021mxu, Fleischhack:2021mhc} aim to search for the final explosions expected to occur at the very end of a BH's life.

Detecting Hawking radiation would not only constitute a powerful probe of quantum field theory in curved space-time but also allow one to measure the properties of a BH, which according to general relativity are simply its mass $M$ and angular momentum $J$ \footnote{Note that, as it evaporates, a BH loses electric charge much faster than mass and spin \cite{Gibbons:1975kk}.}. This is relevant not only for testing general relativity and the cosmological scenarios leading to PBH formation but also for particle physics, since the evolution of both black hole mass and spin through Hawking emission reflects the particle mass spectrum. A BH emits all particle species with mass $m\lesssim T_H$, since for higher mass emission is suppressed by exponential Boltzmann factors $\sim e^{-m/T_H}$. This means, for instance, that observing the final minutes of a PBH's lifetime, when its Hawking temperature reaches values above the TeV scale ($M\lesssim 10^7$ kg), one could potentially probe the existence of new particles beyond the reach of the LHC \cite{Baker:2021btk, Baker:2022rkn,Papanikolaou:2023crz}.

In addition, a BH is not a perfect black body as a result of the non-trivial potential probed by quantum fields in the BH's vicinity. In particular, each quantum field mode has to cross a potential barrier as it propagates away from the BH horizon, the properties of which depend on the BH mass and spin, as well as on the frequency and angular momentum carried by each field mode. As a result, the emission spectrum depends on the intrinsic spin of each field, which also affects the amount of energy and angular momentum removed from the BH by each field quanta during the evaporation process. 

It is well known that, if one assumes only the particle spectrum of the Standard Model (SM) of particle physics and a spin-2 massless graviton, with no yet undiscovered additional degrees of freedom, one finds that a BH loses angular momentum much faster than it loses mass. For instance, for $\tilde{a}\ll 1$ and a Hawking temperature above the electroweak scale ($M\lesssim 10^8$ kg) one obtains $\dot{J}/J\simeq 8 \dot{M}/M$. 
In addition, within the standard paradigm for PBH formation, where the latter results from the gravitational collapse of large overdensities in the radiation-dominated era, PBHs are born with low spins, at or below the percent level \cite{Mirbabayi:2019uph, DeLuca:2019buf}. 

For these reasons, comparatively few works in the literature have so far considered the role of spin in the evolution and cosmological impact of PBHs \cite{Rosa:2017ury,Arbey:2019jmj, Arbey:2019vqx, Hooper:2020evu,Arbey:2020yzj, Ferraz:2020zgi, Cheek:2021odj, Cheek:2022dbx, Cheek:2022mmy, Bernal:2022oha, March-Russell:2022zll, Branco:2023frw,Arbey:2021ysg}, although there are several models in which PBHs may be born or acquire a non-negligible spin parameter.

Firstly, there are cosmological scenarios where PBHs may exhibit large natal spins, even close to extremality ($\tilde{a}\sim 1$) if they are e.g. born in an early matter-dominated epoch \cite{Harada:2017fjm}\footnote{Note that in some scenarios accretion may decrease the PBH spin \cite{deJong:2023gsx}.} or result from the collapse of domain-walls \cite{Eroshenko:2021sez} or scalar field fragmentation \cite{Cotner:2019ykd}. In this case $\dot{J}/J\simeq 2.6 \dot{M}/M$ (within the SM), so one may hope to detect PBHs in the early stages of their evaporation process still exhibiting a substantial spin. 

Secondly, in theoretically well-motivated scenarios with a large number of light scalar species such as the string axiverse \cite{Arvanitaki:2009fg}, PBHs may even develop spin parameters $\tilde{a}< 0.555$ through evaporation \cite{Calza:2021czr}, since emission of scalar quanta with $l=0$ does not change $J$ while decreasing $M$, therefore increasing $\tilde{a}$ \cite{Chambers:1997ai, Taylor:1998dk}. Note that in the SM only the Higgs doublet fields (and pions below the QCD confinement temperature) exhibit this property, but their contribution is overwhelmed by the effects of all other particle species with non-zero spin (fermions and gauge bosons), which necessarily decrease a BH's angular momentum $J$\footnote{Note that for a field of spin $s$ emission occurs only in $l\geq s$ modes.}. Hence, depending on the number of light axion species and natal spin, light PBHs with $M\lesssim 10^{12}$ kg may thus exhibit a non-trivial mass-spin distribution that reflects the underlying particle physics (see also \cite{Calza:2023rjt} for the potential effects of axion superradiant instabilities in this context). 

Devising methodologies to precisely determine a PBH mass and spin is thus an important goal in terms of particle physics, gravity, and cosmology. In our first paper on this topic \cite{Calza:2022ljw}, we began this endeavor by showing that the energies and emission rates of particular features in the electromagnetic Hawking emission spectrum (including both primary and secondary components) can be used to infer both the PBH mass and spin independently of the distance between the latter and the Earth. We focused our discussion on PBHs with spin parameters $0.1\lesssim \tilde{a}\lesssim 0.6$, since this is the range of spins developed by evaporating PBHs in the string axiverse scenario. 

In this second work, we now focus on PBHs with $\tilde{a}\gtrsim 0.6$, for which new features appear in the primary photon emission spectrum as a result of the enhanced angular momentum. As discussed above, values of $\tilde{a}$ in this range would correspond to PBHs born with large spins and that are sufficiently heavy so as not to have lost a significant amount of angular momentum until the present day, assuming they were born in the early Universe (see e.g.~\cite{Picker:2023ybp} for possible scenarios of light BH formation at late times). Note that, even in the axiverse scenario with an arbitrarily large number of light scalar species, a PBH loses spin if it starts with $\tilde{a}>0.555$, since in this regime scalar emission is dominated by non-spherical ($l\geq 1$) modes, which decrease the BH spin.

As in our previous work, our goal is to find observable quantities that are independent of the unknown PBH-Earth distance, although the latter could potentially be determined through parallax techniques if a PBH is sufficiently close \cite{Calza:2021czr}. This is the case of the energy of particular features in the emission spectrum and ratios between the corresponding emission rates. An objective is also to find observables depending exclusively on the PBH spin, or at least a methodology that can clearly disentangle the effects of the PBH mass and spin on the spectrum, such that both properties can be accurately measured. In our previous work the proposed method relied on features of both the primary and secondary photon emission spectrum, the latter corresponding to low energy photons emitted by charged particles in the primary Hawking spectrum as they propagate away from the BH horizon, as well as e.g. decay of primary particles like neutral pions. Although secondary photon emission is more intense than the more energetic primary emission (and hence easier to detect), it has in some regimes a non-trivial shape from which it is hard to extract the BH properties (see \cite{Calza:2021czr} for details). The computation of the secondary spectrum also relies on high-energy physics numerical codes with a limited range of validity, so that there is a non-negligible theoretical uncertainty.

As we will discuss, for large values of the PBH spin parameter $\tilde{a}\gtrsim 0.6$ the primary spectrum exhibits a multi-peak structure, each emission peak corresponding to photons emitted with $l=m=1, 2, 3,\ldots$. Since the emission rate decreases with $l$, one may at least hope to detect the first two emission peaks, corresponding to the $l=m=1$ and $l=m=2$ modes. We will then show that the properties of these two emission peaks yield a robust method to determine the PBH spin and mass.

This work is organized as follows. In the next section, we review the main aspects of the computation of the primary Hawking emission spectrum. In section 3 we discuss our proposed methodology, first based on the total emission rates and later taking into account the effects of anisotropic emission. We summarize our results and discuss prospects for future work in section 4. We also include an appendix discussing a few technical aspects of our calculation, and in particular, validating our proposed methodology by taking into account the secondary component of the Hawking spectrum.

\section{Hawking spectrum}

The differential Hawking emission rate for each particle species $i$ of spin $s$ is given by \cite{Hawking:1975vcx, Page:1976df, Page:1976ki}:
\begin{equation}\label{prim}
{d^2N_{P, i}\over dt dE_i}={1\over 2\pi}\sum_{l,m}{\Gamma^s_{l,m}\over e^{2\pi k/\kappa}\pm 1}~,
\end{equation}
where $\Omega$ is the angular velocity at the BH horizon, at $r=r_+$, $\Gamma^s_{l,m}$ are the transmission coefficients or ``gray-body'' factors encoding the deviations from a black-body spectrum for each $(l,m)$-mode in a spheroidal wave decomposition, $k=\omega-m\Omega$ with $\omega=E_i$ is the mode frequency (in natural units), and $\kappa=\sqrt{1-\tilde{a}^2}/2r_+$ is the surface gravity of a Kerr BH. Note that the plus (minus) sign in the denominator corresponds to fermions (bosons), and that we use the subscript `P' to indicate primary emission, i.e.~particles directly emitted by the BH through the Hawking mechanism.

The non-trivial part of determining the differential emission rate resides in computing the gray-body factors, which are determined by the solutions of the radial Teukolsky equation \cite{Teukolsky:1972my, Teukolsky:1973ha, Press:1973zz, Teukolsky:1974yv} - the master equation that governs the dynamics of (massless) test fields in the Kerr geometry, and which follows from the underlying curved space-time field equations (Klein-Gordon, Dirac, Maxwell and perturbed Einstein equations). The transmission coefficient for each mode can then be computed by considering the associated wave scattering problem and solving the Teukolsky equation through numerical methods. We use a shooting method (see e.g.~\cite{Press:1973zz, Rosa:2016bli} for details) to compute $\Gamma^s_{l,m}$ up to $l=4$ for the $s=1$ massless photons. Although the emission rate is formally an infinite sum over all angular modes, we find that this approximation is sufficiently good for our purposes, taking into account the computational time involved.

We have also explicitly computed the primary spectrum of other emitted particles in order to be able to determine the secondary photon spectrum, as discussed in the appendix. For fermions, we include modes up to $l=7/2$ in our calculation, which is sufficiently accurate\footnote{We have obtained a very good agreement between our results and those obtained in \cite{Dong:2015yjs}, the authors of which have kindly provided us their results for the massless spin-1 gray-body factors.}.


\begin{figure}[h]
\centering\includegraphics[width=0.55\columnwidth]{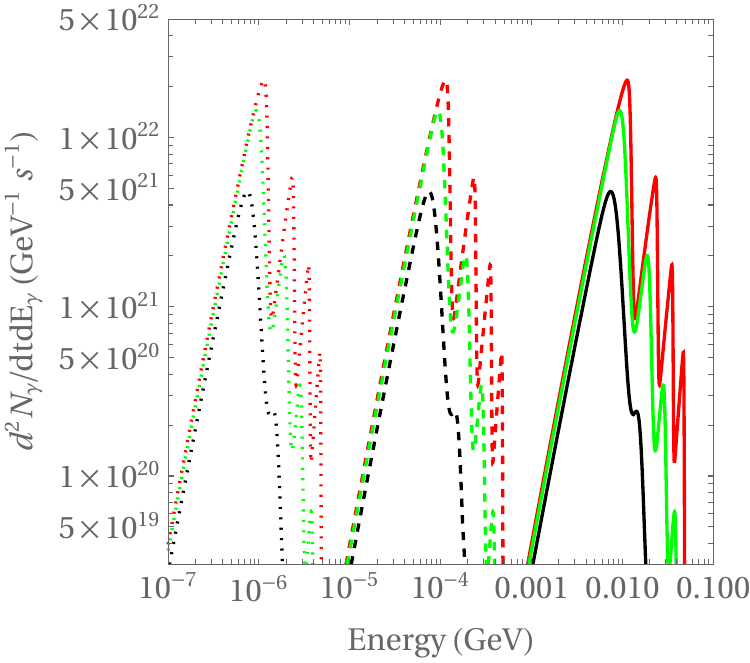}
\caption{Primary photon spectrum for PBHs with $\tilde{a}=0.6, 0.9,  0.99$ (black, green and red, respectively) and $M=10^{13}, 10^{15},  10^{17}$ kg (solid, dashed, and dotted curves, respectively), obtained using Eq.~(\ref{prim}).}\label{fig1}
\end{figure}

Fig.~\ref{fig1} shows examples of the primary Hawking spectrum for different values of the PBH mass and spin $\tilde{a}\geq 0.6$. As one can clearly see, the photon emission rate increases with the PBH spin, a characteristic feature of the emission of particles with non-zero spin, 
while the opposite behavior characterizes scalar emission (see e.g. Fig. 2 in \cite{Arbey:2020yzj}).

\begin{figure}[htbp]
\centering\includegraphics[width=0.55\columnwidth]{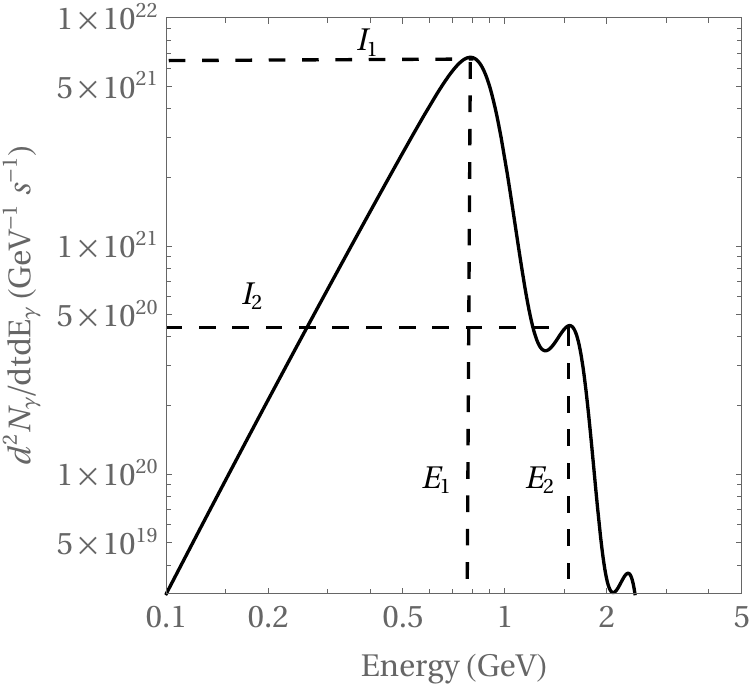}
\caption{Primary photon spectrum for $M=10^{11}$ kg and $\tilde{a}=0.7$, illustrating the energies and emission rates of the first and second peaks used in our analysis.} \label{fig2}
\end{figure}

Independently on the PBH mass, if $\tilde a \gtrsim 0.6$ it is possible to identify in Fig.~\ref{fig1} a primary emission peak and a series of higher energy peaks of decreasing amplitude. As discussed in the appendix, the first peak receives the dominant contribution from the $l=m=1$ mode, the second from the $l=m=2$ mode, and so on. We will focus on the first two peaks since they yield the largest emission rates and hence are easier to detect. We also see that the second peak ($l=2$) is barely discernible for $\tilde{a}=0.6$, so it cannot be employed for PBH spin determination for smaller values of $\tilde{a}$, for which it is overwhelmed by the first peak ($l=1$). In our analysis in the next section, we label the energy and differential emission rate at each peak as $E_l$ and $I_l$, as illustrated in the example of Fig.~\ref{fig2}.

In the next section, we study how these quantities depend on the BH mass and spin and devise a method to determine the latter by measuring these two peaks. Note that detection of the second peak already constitutes an indication of a PBH with $\tilde{a}\gtrsim 0.6$.


\section{PBH mass and spin determination}

\begin{figure}[htbp]
\includegraphics[width=0.5\columnwidth]{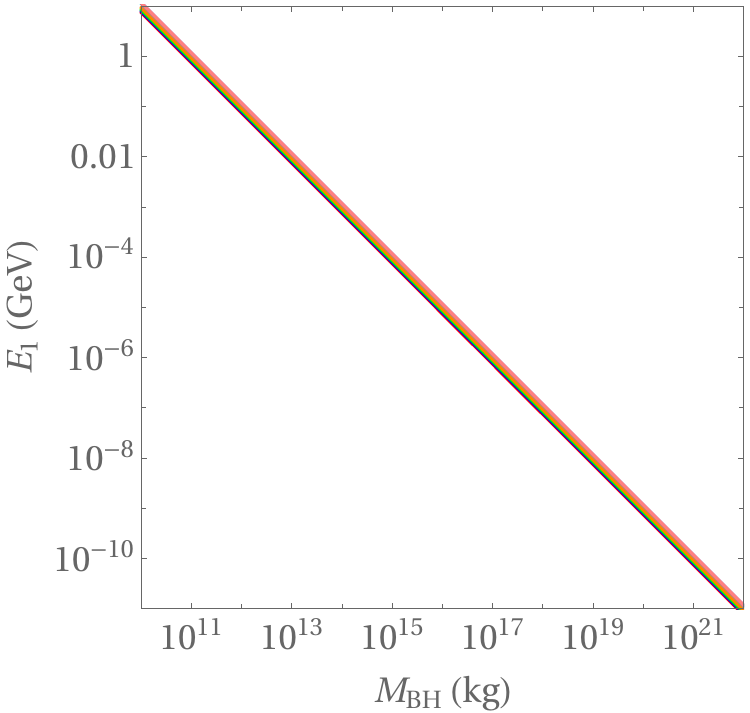}
\includegraphics[width=0.5\columnwidth]{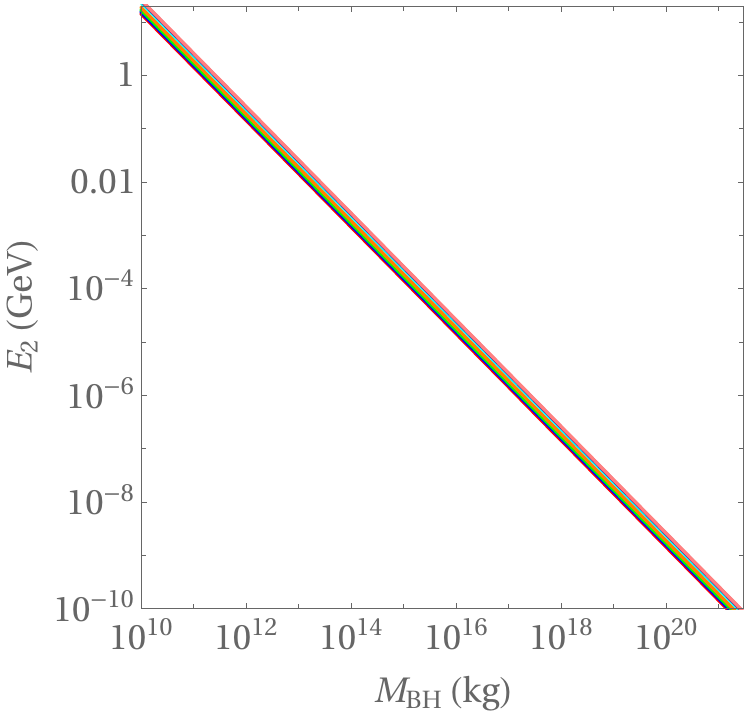}
\caption{Energy of the first ($l=1$, left) and second ($l=2$, right) peaks as a function of the PBH mass for spin parameters $\tilde{a}=0.6, 0.7, 0.8, 0.9,0.99$ (red, blue, green, orange, and cyan, respectively).}\label{fig3}
\end{figure}

In Fig.~\ref{fig3} we show our results for the dependence of the peak energies $E_1$ and $E_2$ on the PBH mass, for different values of dimensionless spin parameter $\tilde a$. 

\begin{figure}[htbp]
\centering\includegraphics[width=0.55\columnwidth]{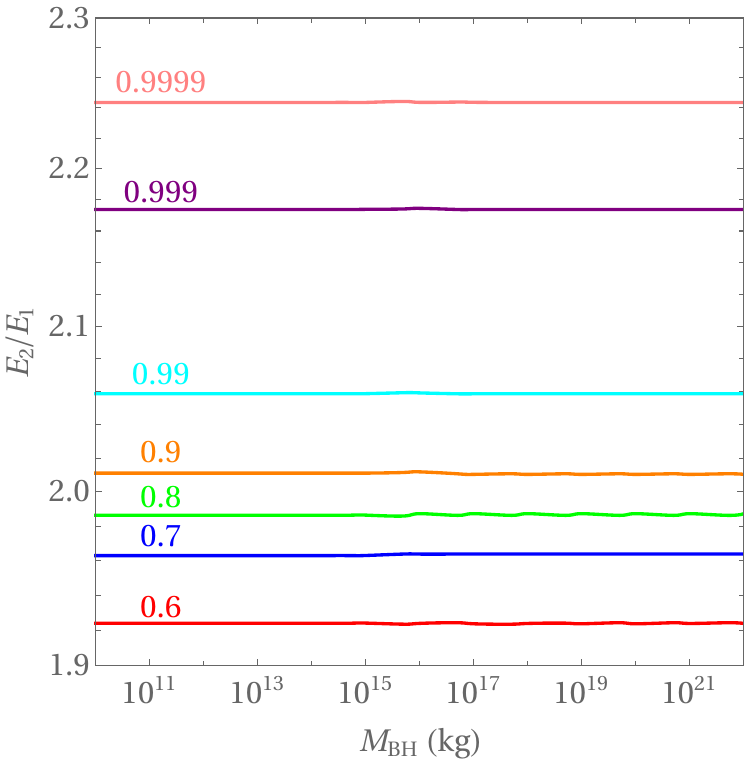}
\caption{Ratio of peak energies $E_2 / E_1$ as a function of the PBH mass for different values of the spin parameter $\tilde{a}$, as labeled.}\label{fig4}
\end{figure}

As this figure clearly illustrates, both peak energies are essentially inversely proportional to the PBH mass, $E_l\propto M^{-1}$, which is simply a consequence of the $M^{-1}$ dependence of the Hawking temperature. The PBH spin dependence is, however, more involved since it is also due to the $\tilde{a}$-dependence of the gray-body factors. For $0.6\leq \tilde{a}< 0.9999$, we obtain in particular:
\begin{equation}
    E_1=\frac{(7.463 - 11.640) \times 10^{10}\ \text{kg}}{M}\ \mathrm{GeV}~, \qquad
    E_2=\frac{(1.436 - 2.611) \times 10^{11}\ \text{kg}}{M}\ \mathrm{GeV}~, \label{eq1P2}
\end{equation}
where the largest numerical factors in the above expressions correspond to near-extremal PBHs. This suggests eliminating the PBH mass dependence by considering the ratio of the two peak energies, $E_2/E_1$, as presented in Fig.~\ref{fig4}. 

This figure shows that this ratio is indeed only a function of the PBH spin, up to small numerical errors in the calculation, and is well described ($<1\%$ deviations) by:
\begin{equation}
    {E_2\over E_1}\simeq 1.761+ 0.284\tilde{a} + \frac{0.000622 \tilde{a}^2}{1-0.997\tilde{a}^2}~, \label{energy_ratio}
\end{equation}

\begin{figure}[htbp]
    \centering
           \centering
    \includegraphics[width=0.7\textwidth]{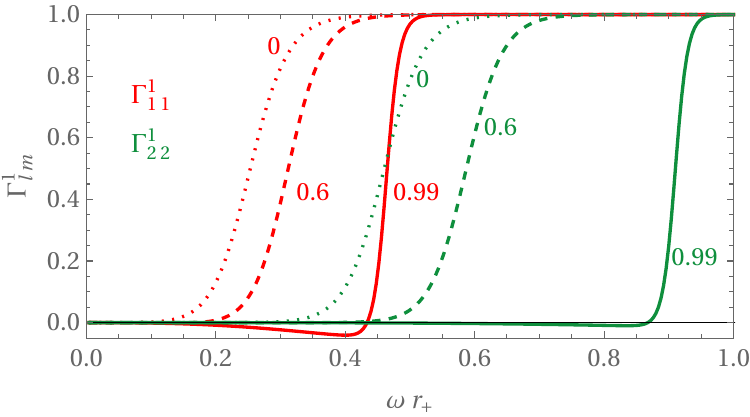}
    \caption{Gray-body factors for the maximally co-rotating modes $l=m=1$ (red) and $l=m=2$ (green) of a massless spin-1 field for $\tilde{a}=0$ (dotted), $\tilde{a}=0.6$ (dashed) and $\tilde{a}=0.99$ (solid).}\label{fig5}
\end{figure}

The fact that this ratio takes values close to 2 is roughly related to the fact that the transmission coefficients, or the gray-body factors in Eq.~(\ref{prim}), exhibit a ``step-like'' behavior, changing rapidly from 0 to 1 at a frequency/energy that grows with the mode's angular momentum $l$, as shown in Fig.~\ref{fig5}. This is associated with the centrifugal potential barrier that wave modes have to tunnel through or overcome as they move away from the BH horizon. In the non-rotating limit (i.e.~a Schwarzschild BH), this barrier has a height $V_\mathrm{max}= 4l(l+1)/27r_+^2$, and only modes with frequency $\omega^2\gtrsim V_\mathrm{max}$ can be fully transmitted. Since the Bose-Einstein distribution in Eq.~(\ref{prim}) is a strictly decreasing function in this limit, the above threshold frequency defines the energy of maximum emission for each mode. Hence, we expect $E_2/E_1\simeq \sqrt{3}\simeq 1.73$ in the non-rotating case, which is close to our numerical fit in Eq.~(\ref{energy_ratio}). Since the PBH spin affects both the gray-body factors and the Bose-Einstein distribution, it is not trivial to obtain an analytical expression for  $\tilde{a}\neq 0$.

Our numerical result shows, nevertheless, that the PBH spin can be determined from a measurement of the ratio between the energies of the second and first peaks of the photon Hawking spectrum. Away from extremality ($0.6\lesssim \tilde{a}\lesssim 0.9$), this ratio has an approximately linear dependence on $\tilde{a}$, and one can easily see that e.g.~a 10$\%$ change in the PBH spin changes the $E_2/E_1$ ratio by only 1-2$\%$. This means that to determine the PBH spin with  $\lesssim 10\%$ uncertainty one needs to measure the peak energies with resolution at or below the percent level. Prospects are better closer to extremal PBH spin values where the $E_2/E_1$ ratio exhibits a more pronounced growth with $\tilde{a}$.

\begin{figure}[htbp]
\centering\includegraphics[width=0.55\columnwidth]{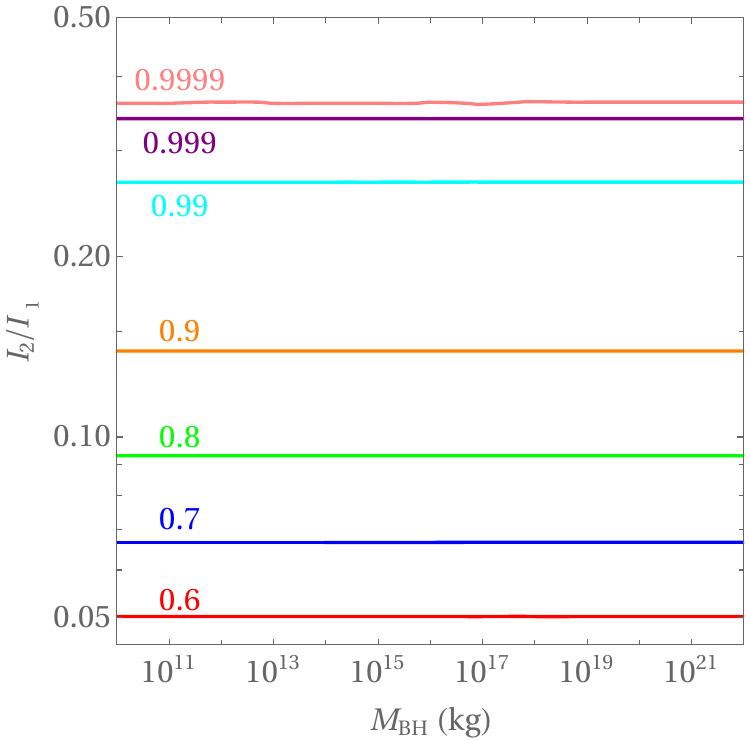}
\caption{Ratio of the peak emission rates $I_2/I_1$ as a function of the PBH mass for different values of the spin parameter $\tilde{a}$, as labeled. }\label{fig6}
\end{figure}

A similar discussion applies to the ratio between the peak emission rates, $I_2/I_1$, since, as shown in Fig.~\ref{fig6}, this is also BH mass-independent, being well described by ($<4\%$):

\begin{equation}
     {I_2\over I_1}\simeq {0.015+0.031\tilde{a}\over 1-0.84\tilde{a}^2}+{0.00241\over \sqrt{1-0.999\tilde{a}^2}}~.
     \label{emission_ratio}
\end{equation}

\begin{figure}[h]
\centering\includegraphics[width=0.55\columnwidth]{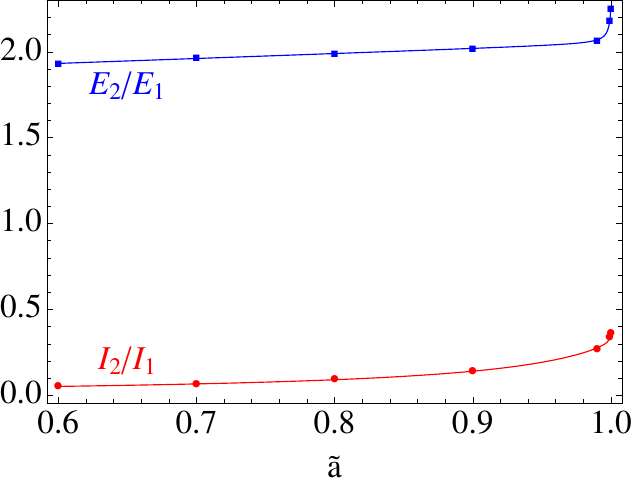}
\caption{Ratios between the energies (blue) and the emission rates (red) of the $l=1$ and $l=2$ peaks in the primary Hawking spectrum as a function of the PBH spin parameter. The points correspond to numerically obtained values while the solid curves yield the approximate analytical expressions given in Eqs.~(\ref{energy_ratio}) and (\ref{emission_ratio}).}\label{fig7}
\end{figure}

We note that this ratio is also independent of the Earth-PBH distance, even though the observed fluxes are. In Fig.~\ref{fig7} we show how the two ratios vary with $\tilde{a}$.

In principle, we would thus have two observable quantities from which one could determine the PBH spin, and therefore a possible consistency check to validate the hypothesis that it corresponds to the Hawking spectrum of a Kerr PBH. However, one must take into account that we cannot really measure the total emission rate but only the flux emitted within a given solid angle along the line of sight. While for non-rotating BHs this is not a problem since the emission is isotropic, this is not the case for spinning Kerr BHs.

\begin{figure}[h!]
    \centering
    \begin{minipage}{0.44\textwidth}
        \centering
    \includegraphics[width=0.99\textwidth]{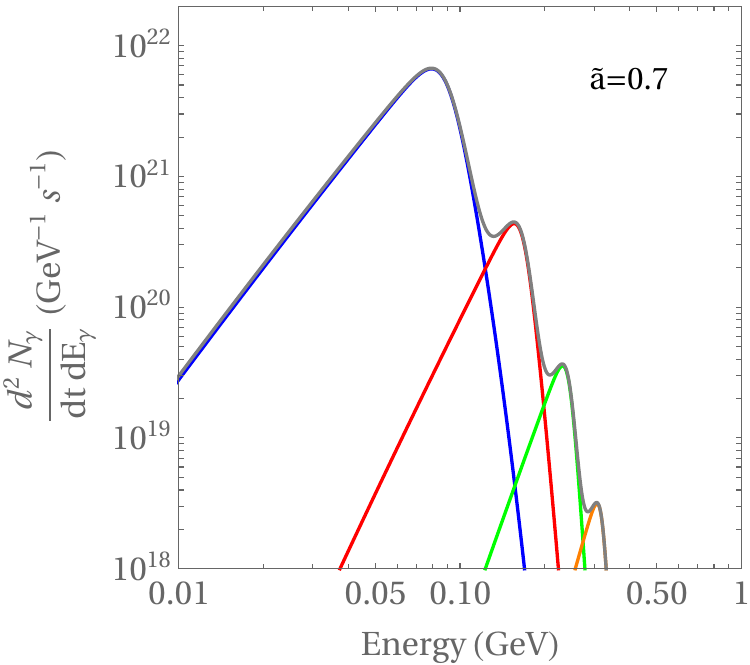}
    \end{minipage}\hfill
    \begin{minipage}{0.44\textwidth}
        \centering
    \includegraphics[width=0.99\textwidth]{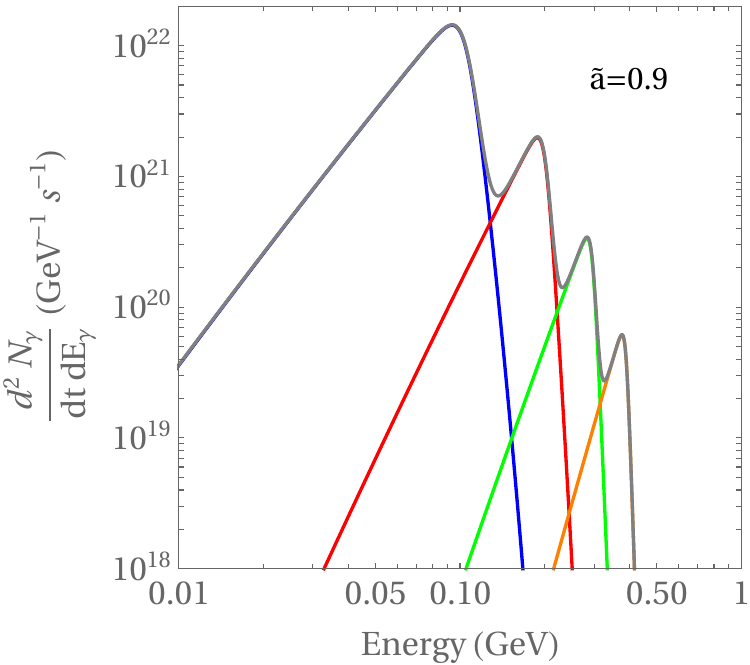}
    \end{minipage}\hfill
    \caption{Superposition of the total primary photon spectrum (gray) and of the contributions of the maximally co-rotating modes $l=m=1,2,3,4$ (blue, red, green, and orange, respectively) for a BH of mass $M=10^{12}$ kg and spin $\tilde{a}= 0.7$ (left) and $\tilde{a}=0.9$ (right).}\label{fig8}
\end{figure}

The angular differential emission rate is given by \cite{Unruh:1973bda,Vilenkin:1978is,Leahy:1979xi,Vilenkin:1979ui,Casals:2012es,Perez-Gonzalez:2023uoi}:
\begin{equation}\label{prim_angular}
{d^2N_{P, i}\over dt dE_id\Omega }={g_s\over{4\pi}}\sum_{l,m}{\Gamma^s_{l,m}\over e^{2\pi k/\kappa}\pm 1} (|S^{-s}_{l,m}(\theta)|^2+|S^{s}_{l,m}(\theta)|^2)~,
\end{equation}
where $S^s_{l,m}(\theta)$ are spheroidal harmonic functions and the angle $\theta$ is measured with respect to the BH rotation axis. As mentioned earlier and shown more explicitly in the appendix, the peaks in the primary spectrum are dominated by the contribution of the maximally co-rotating modes $l=m=1,2,3\ldots$. This is illustrated in Fig.~\ref{fig8}.

Hence, the energy of each peak is independent of the line-of-sight angle $\theta$, since it corresponds essentially to maximizing the contribution to the spectrum of the corresponding $l=m$ mode. This implies that, on the one hand, the ratio $E_2/E_1$ is also independent of $\theta$ and can be used to determine the PBH spin as discussed above.
On the other hand, the intensity of each peak depends on the angle at which we observe the PBH. In particular, the ratio between the second and first peak photon emission fluxes is given by:
\begin{equation}
    {\mathcal{F}_2\over \mathcal{F}_1}={I_2\over I_1}\frac{|S^1_{2,2}(\theta)|^2}{|S^1_{1,1}(\theta)|^2} \simeq {I_2\over I_1}\frac{|Y^1_{2,2}(\theta)|^2}{|Y^1_{1,1}(\theta)|^2} \simeq \frac{5}{3} {I_2\over I_1} \sin^2\theta~,
\end{equation}
where in the last steps we have approximated the spin-1 spheroidal harmonic functions by their spherical counterparts, which holds away from extremality. Note that this implies that the  $l=2$ peak is suppressed relative to the $l=1$ peak when the PBH is observed close to its poles at $\theta=0,\pi$, and in these cases, our proposed methodology cannot be used.

Apart from these special configurations, we conclude that the ratio $E_2/E_1$ can be used to determine the spin of the PBH, while the corresponding flux ratio can be used to infer the inclination of its rotation axis relative to our line-of-sight, $\theta$. 

Optimistically, if a PBH is not too far from the Earth one could detect its proper motion, providing a change in the inclination angle $\theta$ with the characteristic sinusoidal modulation of the ratio of peak fluxes discussed above. For example, PBHs with $\sim 10^{12}$ kg, which have not yet evaporated significantly and may still exhibit large values of $\tilde{a}$, can make up a fraction $f\lesssim 10^{-7}$ of the dark matter. The local dark matter density in our solar neighborhood is around $\rho_{DM}\simeq $0.4 GeV$/\mathrm{cm}^3$, which translates into a typical inter-PBH separation around $\sim 100$ AU so that we could realistically envisage finding PBHs with this mass less than $\sim50$ AU from the Earth. Slightly heavier PBHs with $10^{14}$ kg may in fact make up all the dark matter density in the Universe, which could mean typical separations of just a few AUs, so they could in principle be much closer, as we discuss below. Taking the local velocity of the PBHs to be $\sim 200-300\ \mathrm{km/s}$ relative to the Earth, this implies that each PBH travels a distance of $\sim 40-60$ AU per year, which may translate into an observable proper motion and, in general, to a significant change in inclination.

Measuring the sinusoidal modulation of the peak intensity ratio would then allow one to infer $I_2/I_1$, therefore allowing for a second (redundant) PBH spin determination. This could not only increase the precision of the spin measurement but also serve as a consistency check that the observed spectrum corresponds to an evaporating Kerr BH.

We should note, however, that detecting the primary Hawking emission from light PBHs may require technological advancements beyond the current state-of-the-art technology in terms of gamma-ray telescopes. We may consider the current bounds on the fraction of dark matter in PBHs with $10^{12}-10^{14}$ kg from the contribution of Hawking radiation to the extra-galactic gamma-ray background (see e.g.~\cite{Escriva:2022duf} for a recent review):
\begin{equation}
f(M)<2\times10^{-8} \left({M\over 5\times 10^{11}\ \mathrm{kg}}\right)^{3+\epsilon}~,    
\end{equation}
where $\epsilon=0.1-0.4$ and a monochromatic PBH mass spectrum is assumed. This then yields an upper bound on the number of PBHs of a given mass $M$ that may come within a distance $d$ from the Earth over e.g.~$\Delta t\simeq 10$ yrs, which is a typical lifetime of an experiment:
\begin{equation}
N_{PBH}={f\rho_{DM}\over M} v\pi d^2\Delta t< 1.6\left({v\over 250\ \mathrm{km/s}}\right)\left({d\over 100\ \mathrm{AU}}\right)^2\left({M\over 5\times 10^{11}\ \mathrm{kg}}\right)^{2+\epsilon}~.    
\end{equation}
We may then optimistically assume that the dark matter bounds are saturated (i.e.~that the extra-galactic gamma-ray background receives a significant contribution from PBH evaporation) and compute the distance $d$ for which $N_{PBH}=1$. We can then determine the corresponding photon energy flux at the first peak ($l=1$):
\begin{equation}
\Phi_1\simeq {E_1^2I_1\over 4\pi d^2}~,    
\end{equation}
where we treat the emission as isotropic to estimate its magnitude, keeping in mind the anisotropic effects discussed above. Using our results for the peak energy $E_1$ given in Eq.~(\ref{eq1P2}) in the $\tilde{a}$ range of interest we then have:
\begin{equation}
\Phi_1\lesssim (1.2-3)\times 10^{-8}\left({M\over 5\times 10^{11}\ \mathrm{kg}}\right)^{\epsilon}\left({I_1\over 10^{22}\ \mathrm{GeV}^{-1}\mathrm{s}^{-1}}\right)\ \mathrm{MeV}\mathrm{cm}^{-2}\mathrm{s}^{-1}~,    
\end{equation}
where the spread in numerical factors is related to $\tilde{a}$-dependence of the peak energy $E_1$, with PBHs closer to extremality emitting a larger energy flux. This is maximized for PBHs with $M\sim 10^{14}$ kg that may account for the majority of dark matter and for which we may have $\Phi_1\lesssim 10^{-7}\ \mathrm{MeV}\mathrm{cm}^{-2}\mathrm{s}^{-1}$ (with a weak dependence on the PBH mass in the $10^{12}-10^{14}$ kg mass range). Although this is still roughly an order of magnitude below the expected sensitivity of the planned AMEGO \cite{AMEGO:2019gny, Ray:2021mxu, Fleischhack:2021mhc} and ASTROGAM missions \cite{e-ASTROGAM:2016bph, Tatischeff:2019mun} in the MeV-GeV range, it shows that detecting the primary emission from such PBHs may be possible in the not too distant future. 

Note that, since we have found that $I_2/I_1\simeq 0.05-0.4$ and $E_2/E_1\simeq 1.9-2.2$ in our PBH spin range of interest, we expect $\Phi_2/\Phi_1\simeq 0.2-1.8$. This means that the energy flux from the second peak ($l=2$) may in fact be larger than that of the first peak for PBHs approaching extremality, being at most a factor $5$ smaller than the latter for $\tilde{a}\simeq 0.6$.


\section{Conclusion}

In this work, we have proposed a new method to determine the mass and spin of PBHs from measurements of their primary Hawking radiation spectrum. We have shown, in particular, that the ratio between the energies of the $l=2$ and $l=1$ emission peaks is determined uniquely by the PBH spin parameter $\tilde{a}$, with both energies being inversely proportional to the PBH mass. This thus allows one to determine both Kerr parameters independently of the PBH-Earth distance. The ratio between the corresponding total emission rates is also only a function of $\tilde{a}$, which could yield an additional (redundant) observable to measure the latter. However, since the emission is anisotropic with a distinct angular dependence near each peak, the ratio between the observed fluxes should exhibit a sinusoidal modulation with the line-of-sight. We argued that if a PBH passes sufficiently close to the Earth ($\lesssim 100$ AU), one may be able to detect this modulation from the PBH's proper motion relative to the Solar System.

Since the $l=2$ peak only exceeds the tail of the $l=1$ contribution for $\tilde{a}\gtrsim 0.6$, the proposed method is only applicable in this range. This thus complements the methodology that we have proposed in \cite{Calza:2022ljw} for slowly rotating PBHs, $\tilde{a}\lesssim 0.6$, based on measurements of both the primary and secondary photon spectra. As we show in Appendix B, the secondary photon spectrum does not affect the methodology proposed in this work. We note that large spin parameters $\tilde{a}\gtrsim 0.6$ are only expected for PBHs that have not yet significantly evaporated, even within the context of the string axiverse where light scalar emission can significantly slow down a PBH's spin-down rate (or even spin it up if it is born with low spin values) \cite{Calza:2021czr}. We thus expect the methodology proposed in this work to be applicable to PBHs with $M\gtrsim 10^{12}$ kg. From current bounds on the PBH abundance, we showed that PBHs in the $10^{12}-10^{14}$ kg mass range offer the best prospects for detection of primary Hawking emission in the future. We note that this should require at least an order of magnitude improvement in the sensitivity of gamma-ray telescopes compared to those currently being planned, and we hope that this work motivates further technological developments in gamma-ray detection to achieve this.

\section*{Appendix}
\appendix

\section{Contribution of different modes to the Hawking spectrum}

\begin{figure}[h!]
    \centering
    \begin{minipage}{0.44\textwidth}
        \centering
    \includegraphics[width=0.99\textwidth]{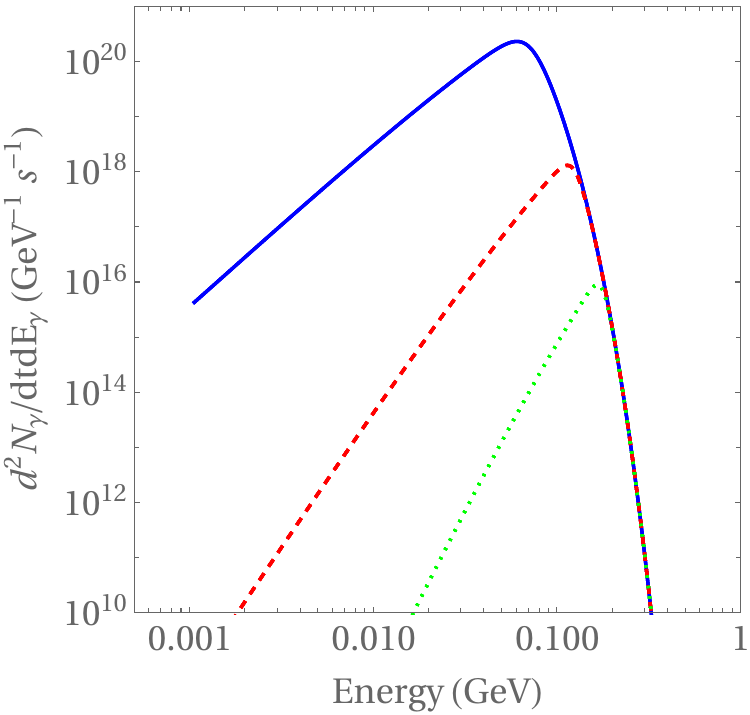}
    \end{minipage}\hfill
    \begin{minipage}{0.44\textwidth}
        \centering
    \includegraphics[width=0.99\textwidth]{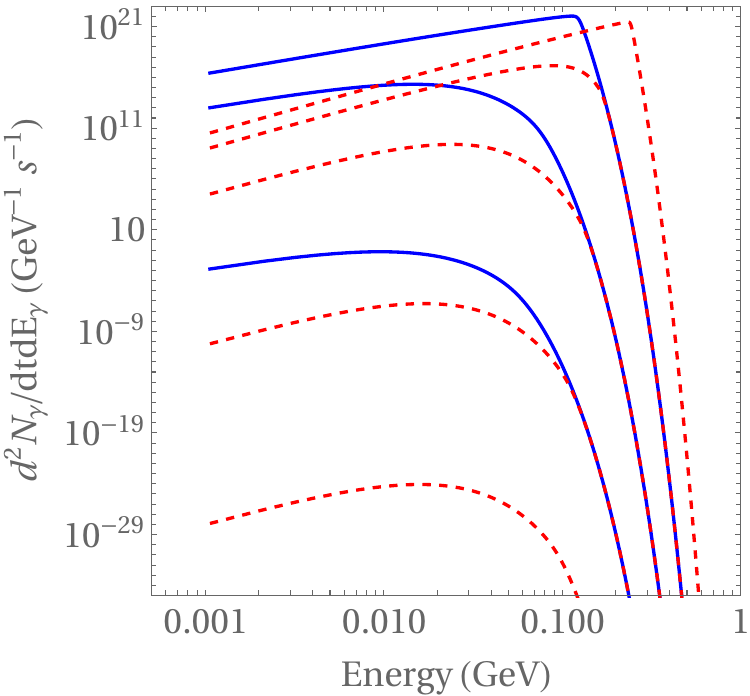}
    \end{minipage}\hfill
    \caption{Contribution of different $(l,m)$ modes to the primary photon spectrum, for a PBH with $M=10^{12}$ kg with $\tilde{a}=0$ (left) and $\tilde{a}=0.99$ (right). Solid blue/dashed red/dotted green curves correspond to $l=1/2/3$ modes, respectively. In the Schwarzschild case modes with the same $l$ are degenerate, while in the near-extremal PBH the curves correspond to $m=-l,\ldots, +l$ from bottom to top.}
    \label{fig9}
\end{figure}

The peaks in the primary photon Hawking spectrum discussed in this work are, as mentioned in the text, dominated by the maximally co-rotating $l=m$ modes for $\tilde{a}>0$. To better understand this, we show in Fig.~\ref{fig9} the contribution of the lowest $(l,m)$ modes, for both limits of a Schwarzschild and a near-extremal BH.

As one can see in this figure, in the non-rotating (Schwarzschild) limit modes with the same $l$ and different $m=-l,\ldots, +l$ yield degenerate contributions to the Hawking emission rate, as should be expected from the spherical symmetry of the space-time. One also sees that the contributions from $l>1$ modes never exceed the dominant contribution from the dipolar ($l=1$) modes. As discussed in the main body of this article, this is simply a consequence of higher-$l$ modes having to cross a higher angular momentum barrier as they propagate away from the BH horizon, so that only modes with a larger energy, and hence a more Boltzmann-suppressed emission, are fully transmitted. This is why only PBHs with a non-zero spin exhibit the multi-peak structure in the primary Hawking spectrum required by our proposed methodology.

For $\tilde{a}>0$, the $m$-degeneracy is broken, and maximally co-rotating modes with $l=m$ give the largest contributions to the spectrum. In fact, as illustrated in Fig.~\ref{fig9}, the emission rate increases with $m$ for a given value of $l$. Superradiant amplification plays a crucial role in breaking this degeneracy since it only occurs for co-rotating modes for which the superradiance condition $\omega<m\Omega$ can be satisfied at low energies/frequencies. For these modes, the exponential Boltzmann factor in Eq.~(\ref{prim}) is below unity, $e^{2\pi (\omega-m\Omega)/\kappa}<1$, and the transmission coefficient $\Gamma_{l,m}^1<0$ (signaling mode amplification, see Fig.~\ref{fig5}), so that both the numerator and denominator of the fraction determining the emission rate are negative. Furthermore, these modes are not Boltzmann-suppressed, therefore yielding significant contributions to the emission spectrum. 

We also observe that the contributions from modes with $l\neq m$ are more suppressed for larger values of $\tilde{a}$, so that in the regime of interest to our spin determination method, $\tilde{a}\gtrsim 0.6$, an accurate expression for the spectrum can be obtained from adding the contributions from only the $l=m$ modes as shown in Fig.~\ref{fig8}. Moreover, since we are mainly interested in the shape of the spectrum near the $l=1$ and $l=2$ peaks, it is sufficient to compute the contributions of modes up to $l=4$ for the level of precision we are interested in\footnote{We note that for other applications such as computing scattering cross sections, one needs to include contributions from higher-$l$ modes.}.

\section{Secondary photon emission spectrum}

An evaporating PBH emits several differently charged particles that radiate photons as they travel away from the PBH. Photons also result from the decay of unstable particles, like neutral pions. These photons are naturally less energetic than those directly emitted but yield nevertheless a very significant contribution to the total photon spectrum at energies smaller than the one of the primary peak. The full spectrum can then be obtained by convoluting the primary emission rate in Eq.~(\ref{prim}) with the number of photons radiated by each charged/unstable primary particle. This is, generically, a non-trivial procedure that has to be performed using numerical tools, particularly in the case of quarks and gluons that hadronize as they move away from the PBH, for Hawking temperatures roughly exceeding $\Lambda_{QCD}$. These reasons often motivate the adoption of toolkits such as BlackHawk \cite{Arbey:2019mbc, Arbey:2020yzj, Arbey:2021yke, Arbey:2021mbl} 
that employ the particle physics codes Hazma and PYTHIA to numerically compute the photon yield of each primary species.

However, since we are mainly interested in PBHs with $M\gtrsim 10^{12}$ kg that may presently exhibit large spin values, $\tilde{a}\gtrsim 0.6$, and for which $T_H\lesssim 10$ MeV, a good approximation to compute the secondary photon spectrum can be obtained by considering only the primary emission of electrons, muons and charged pions and the corresponding final state radiation (FSR), as well as the photon yield from neutral pion decay \cite{Coogan:2020tuf,Agashe:2022jgk}. The total photon spectrum is then given by:
\begin{equation}\label{convo}
{d^2N_{\gamma, tot}\over dt dE_\gamma}={d^2N_{P, \gamma}\over dt dE_\gamma}+
{d^2N_{S, \gamma}^{FSR}\over dt dE_\gamma}+
{d^2N_{S, \gamma}^{decay}\over dt dE_\gamma}~,
\end{equation}
where the subscripts $P$ and $S$ denote the primary and secondary components of the spectrum, respectively. The second term on the right-hand side takes into account the convolution of the primary electron, muon, and charged pion primary spectra as given in Eq.~(\ref{prim}) with the Altarelli-Parisi splitting functions at leading order in the electromagnetic fine-structure constant $\alpha_{EM}$ \cite{Altarelli:1977zs,Chen:2016wkt}:
\begin{equation}
{d^2N_{S, \gamma}^{FSR}\over dt dE_\gamma}= {\sum_{i=e^\pm, \mu^\pm, \pi^\pm}} \int dE_i {d^2N_{P, i}\over dt dE_i} {dN^{FSR}_{i}\over dE_\gamma}, 
\end{equation}
where the contributions from muons and charged pions are sub-leading and:
\begin{equation}
     {dN^{FSR}_{i}\over dE_\gamma}=\frac{\alpha_EM}{\pi Q_i}P_{i \rightarrow i\gamma}(x) \log \left(\frac{1-x}{\mu^2_i}-1\right),
\end{equation}
\begin{equation}
     P_{i \rightarrow i\gamma}(x)= \begin{cases}
  \frac{1+(1-x)^2}{x} & \text{for } i= e^\pm\text{, }\mu^\pm \\
  \frac{2(1-x)}{x} &  \text{for } i=\pi^\pm
  \end{cases},
\end{equation}
with $x= E_\gamma/E_i$ and $\mu_i=m_i/2E_i$. The third term on the right-hand side of Eq.~(\ref{convo}) corresponds to the contribution from neutral pion decays into photon pairs and reads:
\begin{equation}
{d^2N_{S, \gamma}^{decay}\over dt dE_\gamma}=  2\int dE_\pi {d^2N_{P, \pi}\over dt dE_\pi} {dN^{decay}_{\pi}\over dE_\gamma}~,
\end{equation}
where 
\begin{equation}
{dN^{decay}_{\pi}\over dE_\gamma}= \frac{\Theta(E_\gamma-E_\pi^-)\Theta(E_\pi^+-E_\gamma)}{E_\pi^+ - E_\pi^-}
\end{equation}
with $E_\pi^\pm = (E_\pi \pm \sqrt{E_\pi^2-m_\pi^2})/2$.

\begin{figure}[h!]
    \centering
    \begin{minipage}{0.44\textwidth}
        \centering
    \includegraphics[width=0.99\textwidth]{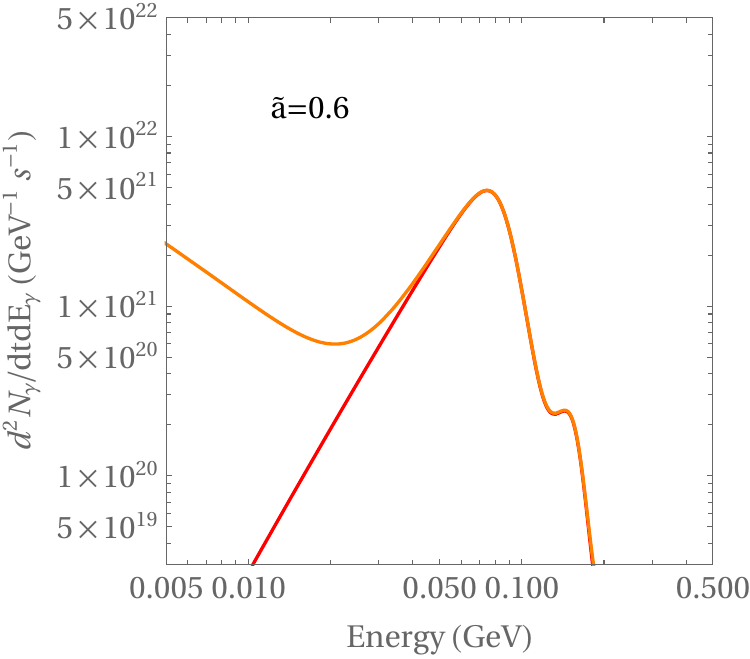}
    \end{minipage}\hfill
    \begin{minipage}{0.44\textwidth}
        \centering
    \includegraphics[width=0.99\textwidth]{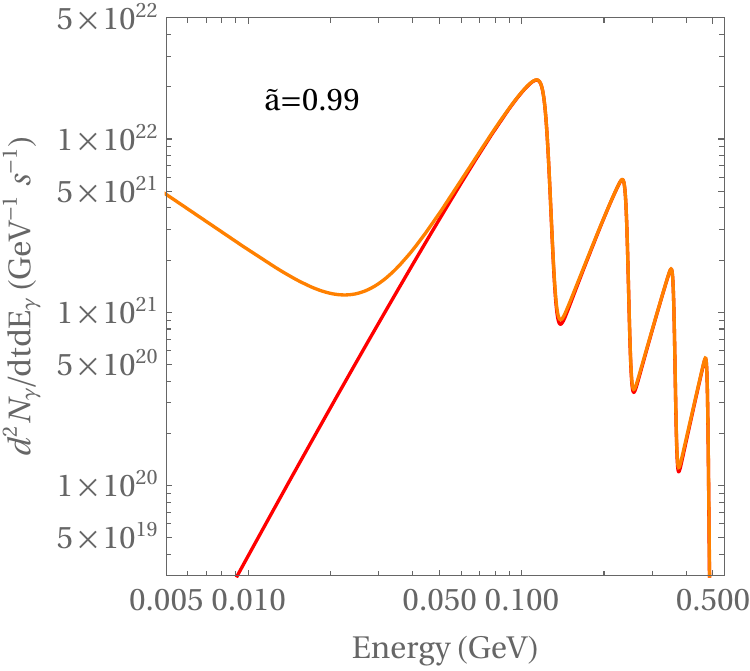}
    \end{minipage}\hfill
    \caption{Primary (red) and total (orange) photon emission spectrum of a PBH with $M=10^{12}$ kg and (left) $\tilde a = 0.6$ and (right) $\tilde a = 0.99$.}\label{fig10}
\end{figure}

In Fig.~\ref{fig10} we show the total photon emission spectrum obtained using this method for a PBH with $M=10^{12}$ kg and two limiting values of the spin parameter relevant to our proposed methodology, alongside the corresponding primary photon component.

This figure shows no substantial change in the shape of the spectrum near the primary emission peaks, particularly the $l=1$ and $l=2$ peaks relevant to our discussion, showing that our analysis neglecting the secondary photon emission spectrum is robust. This should be expected since secondary photons carry only a fraction of the energy of their primary progenitors (electrons, muons, etc), with primary emission for all species peaking at comparable energies $\sim E_1$. Hence, in general, secondary emission is a significant component of the spectrum but only at energies well below the primary photon emission peaks.

\begin{acknowledgments}
We would like to thank Ruifeng Dong, Dejan Stojkovic, William Kinney, and Yuber Perez-Gonzalez for sharing their data for validation of our numerical code, as well as for useful discussions. M.C. is supported by the FCT doctoral grant SFRH/BD/146700/2019. This work was supported by national funds from FCT - Funda\c{c}\~ao para a Ci\^encia e a Tecnologia, I.P., within the project UID/04564/2020 and the grant No.~CERN/FIS-PAR/0027/2021.
\end{acknowledgments}

\end{document}